\documentclass[
 onecolumn,
 11pt, 
 draftcls,
journal
]{IEEEtran}

\usepackage{array}
\usepackage{epsfig}
\usepackage{cite}
\usepackage{amsfonts}
\usepackage{amsmath}
\usepackage{amssymb}
\usepackage{amsthm}
\usepackage{amsxtra}
\usepackage{varioref}
\usepackage{multicol}
\usepackage{float}
\usepackage{rotate}
\usepackage{color}
\usepackage{enumerate}
\usepackage[active]{srcltx} 
\usepackage{mathrsfs}
\usepackage{color}
\usepackage{rotating}
\usepackage{Milancommands}

\usepackage{times}
\usepackage[english]{babel}
\usepackage[utf8]{inputenc}
\usepackage{fontenc}
\usepackage[colorlinks=true,linkcolor=black,citecolor=black]{hyperref}

\title{The Structure of the Realizations of the Causal Information 
Rate-Distortion Function for Markovian Sources:\\
Realizations with Densities} 

 \author{ 
Milan S. Derpich\\
\normalsize
Department of Electronic Engineering\\
Universidad T\'ecnica Federico Santa Mar\'ia.
 }

\begin{document}
\maketitle
\begin{abstract} 
The main purpose of this note is to show that in a realization 
$(\rvax_{1}^{n},\rvay_{1}^{n})$ of the causal \textit{information rate-distortion function} (IRDF) for a 
$\kappa$-th order Markovian
source $\rvax_{1}^{n}$, 
under a single letter sum distortion constraint,
the smallest integer $\ell$ for which
$
  \rvay_{k}
  \leftrightarrow
  \rvay_{1}^{k-1},\rvax_{k-\ell+1}^{k}
  \leftrightarrow
  \rvax_{1}^{k-\ell}
$
holds is $\ell=\kappa$.
This result is derived under the assumption that the sequences $(\rvax_{1}^{n},\rvay_{1}^{n})$ have a joint probability density function.
\end{abstract}

\section{Introduction}
Consider the causal \textit{information rate-distortion function} (IRDF) for a random source 
$\rvax_{1}^{n}=\set{\rvax_{1},\ldots,\rvax_{n}}$,
defined as 
\begin{align}
 R_{c,n}^{it}(D)\eq \frac{1}{n}\inf I(\rvax_{1}^{n};\rvay_{1}^{n})  ,
\end{align}
where the minimization is over all conditional PDFs $f_{\rvay_{1}^{n}|\rvax_{1}^{n}}$ satisfying the
distortion constraint
\begin{align}\label{eq:D_constraint}
\frac{1}{n}\Expe{\sumfromto{i=1}{n}\rho(\rvax_{i},\rvay_{i})}\leq D
\end{align}
and the causality Markov chains 
\begin{align}\label{eq:MC_causality}
 \rvay_{1}^{i}
 \longleftrightarrow
 \rvax_{1}^{i}
 \longleftrightarrow
 \rvay_{i+1}^{n},\fspace i=1,\ldots,n.
\end{align}
If the infimum is achieved by some conditional distribution, the associated pair of sequences $\rvax_{1}^{n},\rvay_{1}^{n}$ is called a realization of $R_{c,n}^{it}(D)$.
Here we assume that such distribution exists and that the corresponding realization has a joint PDF.
This assumption is satisfied if, for example, $\rvax_{1}^{n}$ is Gaussian and $\rho(x,y)=(x-y)^{2}$.

The first purpose of this note is to show that in a realization of the causal IRDF for a 
$\kappa$-th order Markovian
source $\rvax_{1}^{n}$, 
under the average distortion constraint~\eqref{eq:D_constraint},
and supposing that in such realization the sequences have a joint PDF,
it holds that%
\begin{subequations}\label{subeq:first_result}
\begin{align}\label{eq:p_y_given_x_and_y_final_op_F}
  f_{\rvay_{k}|\rvax_{1}^{n},\rvay_{1}^{k-1}}(y_{k}|x_{1}^{n},y_{1}^{k-1})
  &
  =
  \frac
  {
  \expo{- s\rho(x_{k},y_{k}) }
  \breve{F}_{k}(x_{k-\kappa+1}^{k},y_{1}^{k})
  } 
  {
  \int  \expo{- s\rho(x_{k},y_{k}) }
  \breve{F}_{k}(x_{k-\kappa+1}^{k},y_{1}^{k})
  dy_{k}}
\end{align}
where
$f_{\rvax_{1}^{n}}$ is the PDF of $\rvax_{1}^{n}$ and
  \begin{align}\label{subeq:recursion_breve_F}
  \breve{F}_{k}(x_{k-\kappa+1}^{k},y_{1}^{k})
  &
  =
  \expo{ \int  
  \ln \left(\int    \expo{-s\rho(x_{k+1},y_{k+1})} \breve{F}_{k+1}(x_{k-\kappa+2}^{k+1},y_{1}^{k+1}) dy_{k+1} \right)
  f_{\rvax_{k+1}^{n}|\rvax_{k-\kappa+1}^{k}}(x_{k+1}^{n}|x_{k-\kappa+1}^{k})dx_{k+1}^{n} }
\end{align}
\end{subequations}

The expressions given in~\eqref{subeq:first_result} are a special case of the ones given by~\cite[equations (16),(17),(18)]{stacha13} for abstract spaces, where their derivation is not included.
The value of our first result resides in that 
\begin{itemize}
 \item We provide a proof for the validity  of~\eqref{subeq:first_result} (absent in~\cite{stacha13}).
 \item In this proof, 
we pose the causal IRDF optimization problem with $f_{\rvay_{1}^{n}|\rvax_{1}^{n}}$ as the decision variable (instead of the collection $\set{f_{\rvay_{i}|\rvax_{1}^{i},\rvay_{1}^{i-1}} }_{i=1}^{n}$ as would be the case in~\cite{stacha13} for probability measures having an associated PDF).
Accordingly,  we impose an explicit causality constraint on $f_{\rvay_{1}^{n}|\rvax_{1}^{n}}$, instead of enforcing causality structurally by restricting 
$f_{\rvay_{1}^{n}|\rvax_{1}^{n}}$ to be the product of 
$\set{f_{\rvay_{i}|\rvax_{1}^{i},\rvay_{1}^{i-1}} }_{i=1}^{n}$, as done in~\cite{stacha13,chasta14}.
\end{itemize}

The \textbf{second (and main) goal of this document} is to note that from~\eqref{eq:p_y_given_x_and_y_final_op_F}  it is clear that
\begin{align}
  \rvay_{k}
  \longleftrightarrow
  \rvay_{1}^{k-1},\rvax_{k-\kappa+1}^{k}
  \longleftrightarrow
  \rvax_{1}^{k-\kappa}
\end{align}
holds, and that
\begin{align}\label{eq:MC_false}
  \rvay_{k}
  \longleftrightarrow
  \rvay_{1}^{k-1},\rvax_{k}
  \longleftrightarrow
  \rvax_{1}^{k-1}
\end{align}  
\textbf{does not hold, except for $\kappa=1$.
Crucially,~\eqref{eq:MC_false} does not become true by supposing that the joint PDF of $\rvax_{1}^{k},\rvay_{1}^{k}$ is stationary, thus contradicting~\cite[Remark IV.5]{chasta14} and what is stated in the discussion paragraph at the end of~\cite[Section~V]{stacha13}.}

\section{Proof}
The causal IRDF under the above conditions is yielded by the solution to the following optimization problem:
\begin{subequations}\label{subeq:OPcausality}
\begin{align}
  \text{minimize: } \fspace & I(\rvax_{1}^{n};\rvay_{1}^{n})
  \\
  \text{subject to: } \fspace & 
   \left(\int  f_{\rvay_{1}^{n}|\rvax_{1}^{n}}(y_{1}^{n}|x_{1}^{n})dy_{1}^{n} -1\right)f_{\rvax_{1}^{n}}(x_{1}^{n}) =0,\fspace \forall x_{1}^{n}\label{eq:int_1_constraint}\\
			& \iint  f_{\rvay_{1}^{n}|\rvax_{1}^{n}}(y_{1}^{n}|x_{1}^{n})f_{\rvax_{1}^{n}}(x_{1}^{n})\sumfromto{k=1}{n}\rho(x_{k},y_{k})dy_{1}^{n}dx_{1}^{n} \leq D\\
			& (f_{\rvay_{1}^{k}|\rvax_{1}^{k}}(y_{1}^{k}|x_{1}^{k}) 
			- f_{\rvay_{1}^{k}|\rvax_{1}^{n}}(y_{1}^{k}|x_{1}^{n}))
			f_{\rvax_{1}^{n}}(x_{1}^{n}) =0,\fspace  \forall y_{1}^{k},x_{1}^{n}, \; k=1,\ldots,n.\label{eq:causality_constraint}
\end{align}
\end{subequations}
where the minimization is over the conditional PDF $f_{\rvay_{1}^{n}|\rvax_{1}^{n}}$.
\textbf{Notice that~\eqref{eq:causality_constraint} is an explicit causality constraint} equivalent to~\eqref{eq:MC_causality}.

Let $f'_{\rvay_{1}^{n}|\rvax_{1}^{n}}:\Rl^{n\times n}\to [0,1]$ be any conditional PDF, and define  
\begin{align}
 g_{\rvay_{1}^{n}|\rvax_{1}^{n}}
 &\eq (f'_{\rvay_{1}^{n}|\rvax_{1}^{n}}-f_{\rvay_{1}^{n}|\rvax_{1}^{n}})\label{eq:g_def}
 \\
 g_{\rvay_{1}^{n}}(y_{1}^{n})
 &\eq 
 \int 
 g_{\rvay_{1}^{n}|\rvax_{1}^{n}}(y_{1}^{n}|x_{1}^{n})
 f_{\rvax_{1}^{n}}(x_{1}^{n})dx_{1}^{n}
 \\
 f^{\e}_{\rvay_{1}^{n}|\rvax_{1}^{n}} &\eq f_{\rvay_{1}^{n}|\rvax_{1}^{n}} +\e g_{\rvay_{1}^{n}|\rvax_{1}^{n}}
 \\
 f^{\e}_{\rvay_{1}^{n}}(y_{1}^{n}) &\eq 
 \int   f^{\e}_{\rvay_{1}^{n}|\rvax_{1}^{n}}(y_{1}^{n}|x_{1}^{n})f_{\rvax_{1}^{n}}(x_{1}^{n})dx_{1}^{n}
\end{align}
where $\e\in[0,1]$.

Before writing the Lagrangian and taking its Gateaux differential, let us obtain the Gateaux differential of $I(\rvax_{1}^{n};\rvay_{1}^{n})$ in the direction 
$g_{\rvay_{1}^{n}|\rvax_{1}^{n}}$, given by 
\begin{align}
 \frac{d I(\rvax_{1}^{n};\rvay_{1}^{n})}{d\e}
 \Big |_{\e=0}
 &=
 \frac{d }{d\e}
 \left[
 \iint f^{\e}_{\rvay_{1}^{n}|\rvax_{1}^{n}}(y_{1}^{n}|x_{1}^{n})f_{\rvax_{1}^{n}}(x_{1}^{n})
 \ln\left(\frac{f^{\e}_{\rvay_{1}^{n}|\rvax_{1}^{n}}(y_{1}^{n}|x_{1}^{n})}{f^{\e}_{\rvay_{1}^{n}}(y_{1}^{n})} \right)dy_{1}^{n}dx_{1}^{n}
 \right]\Bigg |_{\e=0}
 \\&
 =
 \iint g_{\rvay_{1}^{n}|\rvax_{1}^{n}}(y_{1}^{n}|x_{1}^{n})f_{\rvax_{1}^{n}}(x_{1}^{n})
 \ln\left(\frac{f_{\rvay_{1}^{n}|\rvax_{1}^{n}}(y_{1}^{n}|x_{1}^{n})}{f_{\rvay_{1}^{n}}(y_{1}^{n})} \right)
 dy_{1}^{n}dx_{1}^{n}+
 R
\end{align}
where 
\begin{align}
 R
 &\eq
  \iint  f_{\rvay_{1}^{n}|\rvax_{1}^{n}}(y_{1}^{n}|x_{1}^{n})f_{\rvax_{1}^{n}}(x_{1}^{n})
 \left(
 \frac{g_{\rvay_{1}^{n}|\rvax_{1}^{n}}(y_{1}^{n}|x_{1}^{n})}{f_{\rvay_{1}^{n}|\rvax_{1}^{n}}(y_{1}^{n}|x_{1}^{n}) }
 -
 \frac{g_{\rvay_{1}^{n}}(y_{1}^{n})} {f_{\rvay_{1}^{n}}(y_{1}^{n}) }
 \right)dy_{1}^{n}dx_{1}^{n}
\\&
=
 \iint   g_{\rvay_{1}^{n}|\rvax_{1}^{n}}(y_{1}^{n}|x_{1}^{n})f_{\rvax_{1}^{n}}(x_{1}^{n})dy_{1}^{n}dx_{1}^{n}
 -
 \iint 
 \frac{   f_{\rvay_{1}^{n},\rvax_{1}^{n}}(y_{1}^{n},x_{1}^{n})g_{\rvay_{1}^{n}}(y_{1}^{n})} {f_{\rvay_{1}^{n}}(y_{1}^{n}) }dy_{1}^{n}dx_{1}^{n}
 \\&
=
 \int g_{\rvay_{1}^{n}}(y_{1}^{n})dy_{1}^{n}
 -
 \int
 \frac{g_{\rvay_{1}^{n}}(y_{1}^{n}) } {f_{\rvay_{1}^{n}}(y_{1}^{n}) }
 \left(\int 
 f_{\rvay_{1}^{n},\rvax_{1}^{n}}(y_{1}^{n},x_{1}^{n})dx_{1}^{n}
 \right)dy_{1}^{n}
 \\&
 =0
\end{align}

On the other hand, for each $i=1,\ldots,n$, the causality constraint~\eqref{eq:causality_constraint} appears in the Lagrangian as 
\begin{align}
  \iint & \lambda_{i}(x_{1}^{n},y_{1}^{i})\left[f_{\rvay_{1}^{i}|\rvax_{1}^{i}}(y_{1}^{i}|x_{1}^{i}) 
							  - f_{\rvay_{1}^{i}|\rvax_{1}^{n}}(y_{1}^{i}|x_{1}^{n})\right]
							  f_{\rvax_{1}^{n}}(x_{1}^{n})dy_{1}^{i}dx_{1}^{n}
\\
=&
\iint 
\lambda_{i}(x_{1}^{n},y_{1}^{i})
\left(
\int   \left[f_{\rvay_{1}^{n}|\rvax_{1}^{i}}(y_{1}^{n}|x_{1}^{i}) 
							  - f_{\rvay_{1}^{n}|\rvax_{1}^{n}}(y_{1}^{n}|x_{1}^{n})\right]dy_{i+1}^{n}\right)
							  f_{\rvax_{1}^{n}}(x_{1}^{n})dy_{1}^{i}dx_{1}^{n}
\\
=&
\int 
\left(
\int\lambda_{i}(x_{1}^{n},y_{1}^{i}) f_{\rvay_{1}^{n}|\rvax_{1}^{i}}(y_{1}^{n}|x_{1}^{i}) 
f_{\rvax_{1}^{n}}(x_{1}^{n})dx_{1}^{n}
							  -
\int \lambda_{i}(x_{1}^{n},y_{1}^{i})	f_{\rvay_{1}^{n}|\rvax_{1}^{n}}(y_{1}^{n}|x_{1}^{n})
f_{\rvax_{1}^{n}}(x_{1}^{n})dx_{1}^{n}
\right)dy_{1}^{n}
\label{eq:shankar}
\end{align}
It will be convenient to manipulate this expression so as to give it a structure similar to the other terms in the Lagrangian.
For this purpose, notice that 
\begin{align}
\int\lambda_{i}(x_{1}^{n},y_{1}^{i}) f_{\rvay_{1}^{n}|\rvax_{1}^{i}}(y_{1}^{n}|x_{1}^{i}) 
&f_{\rvax_{1}^{n}}(x_{1}^{n})dx_{1}^{n}
  \\&
  =
\int \lambda_{i}(x_{1}^{n},y_{1}^{i}) f_{\rvay_{1}^{n},\rvax_{1}^{i}}(y_{1}^{n},x_{1}^{i}) 
f_{\rvax_{i+1}^{n}|\rvax_{1}^{i}}(x_{i+1}^{n}|x_{1}^{i})dx_{1}^{n}
  \\&
  =
\int f_{\rvay_{1}^{n},\rvax_{1}^{i}}(y_{1}^{n},x_{1}^{i}) 
\left(
\int\lambda_{i}(x_{1}^{n},y_{1}^{i})  f_{\rvax_{i+1}^{n}|\rvax_{1}^{i}}(x_{i+1}^{n}|x_{1}^{i})dx_{i+1}^{n}\right)dx_{1}^{i}
  \\&
  =
\int f_{\rvay_{1}^{n},\rvax_{1}^{i}}(y_{1}^{n},x_{1}^{i}) 
\bar{\lambda}(x_{1}^{i},y_{1}^{i})dx_{1}^{i}
  \\&
  =
\int
\left(\int f_{\rvay_{1}^{n},\rvax_{1}^{n}}(y_{1}^{n},x_{1}^{n}) dx_{i+1}^{n}\right)
\bar{\lambda}(x_{1}^{i},y_{1}^{i})dx_{1}^{i}
\\&
=
\int  f_{\rvay_{1}^{n}|\rvax_{1}^{n}}(y_{1}^{n}|x_{1}^{n}) 
f_{\rvax_{1}^{n}}(x_{1}^{n})
\bar{\lambda}(x_{1}^{i},y_{1}^{i})dx_{1}^{n}
\end{align}
where 
\begin{align}\label{eq:barlambda_def}
  \bar{\lambda}_{i}(x_{1}^{i},y_{1}^{i})
  \eq\int \lambda_{i}(x_{1}^{n},y_{1}^{i})  f_{\rvax_{i+1}^{n}|\rvax_{1}^{i}}(x_{i+1}^{n}|x_{1}^{i})dx_{i+1}^{n}
  ,\fspace i=1,\ldots,n.
\end{align}
Substituting this into~\eqref{eq:shankar}
we obtain
\begin{align}
  \int & \lambda_{i}(x_{1}^{n},y_{1}^{i})(f_{\rvay_{1}^{i}|\rvax_{1}^{i}}(y_{1}^{i}|x_{1}^{i}) 
							  - f_{\rvay_{1}^{i}|\rvax_{1}^{n}}(y_{1}^{i}|x_{1}^{n}))
							  f_{\rvax_{1}^{n}}(x_{1}^{n})dy_{1}^{i}dx_{1}^{n}
							  \\&
=
\int
\left(
  \bar{\lambda}_{i}(x_{1}^{i},y_{1}^{i})
  -
    \lambda_{i}(x_{1}^{n},y_{1}^{i})
    \right)
f_{\rvay_{1}^{n}|\rvax_{1}^{n}}(y_{1}^{n}|x_{1}^{n})
f_{\rvax_{1}^{n}}(x_{1}^{n})dy_{1}^{n}dx_{1}^{n}
\end{align}

We can now write the Lagrangian associated with optimization problem~\eqref{subeq:OPcausality} as 
\begin{align}
\Lsp(f_{\rvay_{1}^{n}|\rvax_{1}^{n}})&
\eq 
I(\rvax_{1}^{n};\rvay_{1}^{n})
+
\int\eta(x_{1}^{n})\left(\int f_{\rvay_{1}^{n}|\rvax_{1}^{n}}(y_{1}^{n}|x_{1}^{n})dy_{1}^{n} -1\right)f_{\rvax_{1}^{n}}(x_{1}^{n})dx_{1}^{n} 
\\&
+
s\left(
\int f_{\rvay_{1}^{n}|\rvax_{1}^{n}}(y_{1}^{n}|x_{1}^{n})f_{\rvax_{1}^{n}}(x_{1}^{n})
\left(\sumfromto{i=1}{n}\rho(x_{i},y_{i})\right)dx_{1}^{n}dy_{1}^{n}- D\right)
\\&
+
\Sumfromto{i=1}{n}
\int 
\left(
  \bar{\lambda}_{i}(x_{1}^{i},y_{1}^{i})
  -
    \lambda_{i}(x_{1}^{n},y_{1}^{i})
    \right)
f_{\rvay_{1}^{n}|\rvax_{1}^{n}}(y_{1}^{n}|x_{1}^{n})
f_{\rvax_{1}^{n}}(x_{1}^{n})dy_{1}^{n}dx_{1}^{n}
\end{align}
From the theory of Lagrangian  optimization on vector spaces~\cite{luenbe69}, 
$f_{\rvay_{1}^{n}|\rvax_{1}^{n}}$  is a 
solution to Optimization Problem~\eqref{subeq:OPcausality} only if
\begin{align}
0&=
 \frac{d}{d\e}
 \Lsp(f^{\e}_{\rvay_{1}^{n}|\rvax_{1}^{n}})\Big|_{\e=0}
 \\&
 =
\Sumover{y_{1}^{n},x_{1}^{n}} 
\Bigg[ 
 \ln\left(\frac{f_{\rvay_{1}^{n}|\rvax_{1}^{n}}(y_{1}^{n}|x_{1}^{n})}{f_{\rvay_{1}^{n}}(y_{1}^{n})} \right)
+
\eta(x_{1}^{n}) 
  +
 \sumfromto{i=1}{n}
 \left(
  s\rho(x_{i},y_{i})
 +
  \bar{\lambda}_{i}(x_{1}^{i},y_{1}^{i})
  -
    \lambda_{i}(x_{1}^{n},y_{1}^{i})
    \right)
 \Bigg]\nonumber
 \\&
 \fspace \fspace\fspace\fspace\fspace\fspace\fspace\fspace\fspace\fspace\fspace
  \fspace \fspace\fspace\fspace\fspace\fspace
 \times 
 g_{\rvay_{1}^{n}|\rvax_{1}^{n}}(y_{1}^{n}|x_{1}^{n})
f_{\rvax_{1}^{n}}(x_{1}^{n})
 \end{align}
 for every function $g_{\rvay_{1}^{n}|\rvax_{1}^{n}}$ as defined in~\eqref{eq:g_def}, i.e., for every conditional PDF $f'_{\rvay_{1}^{n}|\rvax_{1}^{n}}$.
This holds if and only if for every $x_{1}^{n},y_{1}^{n}$:
\begin{align}
 \ln\left(\frac{f_{\rvay_{1}^{n}|\rvax_{1}^{n}}(y_{1}^{n}|x_{1}^{n})}{f_{\rvay_{1}^{n}}(y_{1}^{n})} \right)
&=
-\eta(x_{1}^{n}) 
  -
 \sumfromto{i=1}{n}
 \left(
  s\rho(x_{i},y_{i})
 +
  \bar{\lambda}_{i}(x_{1}^{i},y_{1}^{i})
  -
    \lambda_{i}(x_{1}^{n},y_{1}^{i})
    \right)
     \\
     \iff 
     f_{\rvay_{1}^{n}|\rvax_{1}^{n}}(y_{1}^{n}|x_{1}^{n})
    &
    =
    \expo{-\eta(x_{1}^{n}) 
  -
  \sumfromto{i=1}{n}
 \left(
  s\rho(x_{i},y_{i})
 +
  \bar{\lambda}_{i}(x_{1}^{i},y_{1}^{i})
  -
    \lambda_{i}(x_{1}^{n},y_{1}^{i})
    \right)
 }f_{\rvay_{1}^{n}}(y_{1}^{n})
 \end{align}
The Lagrange multiplier function $\eta(x_{1}^{n})$ must enforce the constraint~\eqref{eq:int_1_constraint}.
Hence, 
\begin{align}\label{eq:lamaestra}
 f_{\rvay_{1}^{n}|\rvax_{1}^{n}}(y_{1}^{n}|x_{1}^{n})
    &
    =
  \frac{\expo{ -  \sumfromto{i=1}{n} \left(  s\rho(x_{i},y_{i}) +  \bar{\lambda}_{i}(x_{1}^{i},y_{1}^{i})  -    \lambda_{i}(x_{1}^{n},y_{1}^{i})    \right)}
    f_{\rvay_{1}^{n}}(y_{1}^{n})
  }
  {K_{1}(x_{1}^{n})},
\end{align}
where 
\begin{align}\label{eq:K_j_def}
 K_{1}(x_{1}^{n}) &\eq 
 \int 
 \expo{ -\sumfromto{i=1}{n}\left(s\rho(x_{i},y_{i}) + \bar{\lambda}_{i}(x_{1}^{i},y_{1}^{i})-\lambda_{i}(x_{1}^{n},y_{1}^{i})\right)} 
 f_{\rvay_{1}^{n}}(y_{1}^{n})dy_{1}^{n}
 \end{align}
Marginalizing over $y_{k+1}^{n}$ we obtain
\begin{align}
  f_{\rvay_{1}^{k}|\rvax_{1}^{n}}(y_{1}^{k}|x_{1}^{n})
    &
    =
  \frac{
  \expo{ -  \sumfromto{i=1}{k} (s\rho(x_{i},y_{i}) +  \bar{\lambda}_{i}(x_{1}^{i},y_{1}^{i})-\lambda_{i}(x_{1}^{n},y_{1}^{i})    )}
  \int 
  \expo{ -  \sumfromto{i=k+1}{n} \left(  s\rho(x_{i},y_{i}) +  \bar{\lambda}_{i}(x_{1}^{i},y_{1}^{i})  -    \lambda_{i}(x_{1}^{n},y_{1}^{i})    \right)} f_{\rvay_{1}^{n}}(y_{1}^{n})dy_{k+1}^{n}
  }
  {K_{1}(x_{1}^{n})}
\end{align}
Using Bayes' rule we can write
\begin{align}
   f_{\rvay_{k}|\rvax_{1}^{n},y_{1}^{k-1}}(y_{k}|x_{1}^{n},y_{1}^{k-1})
    &
    =
    \frac{f_{\rvay_{1}^{k}|\rvax_{1}^{n}}(y_{1}^{k}|x_{1}^{n})}{f_{\rvay_{1}^{k-1}|\rvax_{1}^{n}}(y_{1}^{k-1}|x_{1}^{n})}
  \\&
  =
  \frac{
  \expo{- s\rho(x_{k},y_{k})  }
  F_{k}(x_{1}^{n},y_{1}^{k})
  }
  {
  \int \expo{- s\rho(x_{k},y_{k}) } 
  F_{k}(x_{1}^{n},y_{1}^{k})dy_{k}
  }
   \label{eq:simple_op_F}
  \end{align}
where 
\begin{align}
  F_{k}(x_{1}^{n},y_{1}^{k})
  &\eq
  \expo{-(\bar{\lambda}_{k}(x_{1}^{k},y_{1}^{k})-\lambda_{k}(x_{1}^{n},y_{1}^{k}))}
  \int \expo{ -  \sumfromto{i=k+1}{n} \left(  s\rho(x_{i},y_{i}) +  \bar{\lambda}_{i}(x_{1}^{i},y_{1}^{i})  -    \lambda_{i}(x_{1}^{n},y_{1}^{i})    \right)} f_{\rvay_{1}^{n}}(y_{1}^{n})dy_{k+1}^{n}
\end{align}
These functions can be written recursively as 
\begin{subequations}\label{subeq_recursion_F}
  \begin{align}
 F_{n}(y_{1}^{n})
&= f_{\rvay_{1}^{n}}(y_{1}^{n})\label{eq:Fn_def}
\\
  F_{k}(x_{1}^{n},y_{1}^{k})
  &
  =
  \expo{-(\bar{\lambda}_{k}(x_{1}^{k},y_{1}^{k})-\lambda_{k}(x_{1}^{n},y_{1}^{k}))}
  \int
  \expo{-s\rho(x_{k+1},y_{k+1})} F_{k+1}(x_{1}^{n},y_{1}^{k+1}) dy_{k+1}
  \label{eq:F_k_recursive}
\end{align}
\end{subequations}
In order attain causality in~\eqref{eq:simple_op_F}, the functions $F_{k}(x_{1}^{n},y_{1}^{k})$ must depend only on $x_{1}^{k}$ and $y_{1}^{k}$.
Since for each $k$, the function $F_{k+1}$ does not depend on terms 
$(\bar{\lambda}_{i}(x_{1}^{i},y_{1}^{i})-\lambda_{i}(x_{1}^{n},y_{1}^{i}))$ with $i\leq k$,
the causality constraint is met if and only if we choose 
$(\bar{\lambda}_{i}(x_{1}^{k},y_{1}^{k})-\lambda_{i}(x_{1}^{n},y_{1}^{k}))$
in~\eqref{eq:F_k_recursive}
such that, for each $k=1,\ldots,n$ 
\begin{align}\label{eq:Fk_breve_Fk}
F_{k}(x_{1}^{n},y_{1}^{k})
=
  \expo{-(\bar{\lambda}_{i}(x_{1}^{k},y_{1}^{k})-\lambda_{i}(x_{1}^{n},y_{1}^{k}))}
  \int
  \expo{-s\rho(x_{k+1},y_{k+1})} F_{k+1}(x_{1}^{n},y_{1}^{k+1}) dy_{k+1}
  =
  \breve{F}_{k}(x_{1}^{k},y_{1}^{k})
\end{align}
for some function $\breve{F}_{k}$.

For $k=n$, the causality constraint is satisfied automatically since 
$F_{n}(x_{1}^{n},y_{1}^{n}) =  \breve{F}_{n} (y_{1}^{k})\eq f_{\rvay_{1}^{n}}(y_{1}^{n})$ (see~\eqref{eq:Fn_def}).%
\footnote{
This reflects the fact that there is no need to enforce the causality constraint for $k=n$, since there are no source samples for time $k>n$.}
Suppose now that~\eqref{eq:Fk_breve_Fk} (i.e., causality) is satisfied for $k+1$, for some $k>n$. 
In such case, one can replace 
$F_{k+1}(x_{1}^{n},y_{1}^{k+1})$ in~\eqref{eq:Fk_breve_Fk} by 
$\breve{F}_{k+1}(x_{1}^{k+1},y_{1}^{k+1})$ and, 
defining 
$$
K_{k+1}(x_{1}^{k+1},y_{1}^{k})
\eq
\int  \expo{-s\rho(x_{k+1},y_{k+1})} \breve{F}_{k+1}(x_{1}^{k+1},y_{1}^{k+1}) dy_{k+1},
$$
write~\eqref{eq:Fk_breve_Fk} as
\begin{align}\label{eq:barlambda_minus_lambda}
\bar{\lambda}_{k}(x_{1}^{k},y_{1}^{k})
  -
  \lambda_{k}(x_{1}^{n},y_{1}^{k})
 &=
\ln K_{k+1}(x_{1}^{n},y_{1}^{k})
-
\ln \breve{F}_{k}(x_{1}^{k},y_{1}^{k})
  \end{align}
  Multiplying both sides by $f_{\rvax_{k+1}^{n}|\rvax_{1}^{k}}(x_{k+1}^{n}|x_{1}^{k})$ and 
  integrating over $x_{k+1}^{n}$ we obtain 
  \begin{align}
 0&= \int \left(\bar{\lambda}_{k}(x_{1}^{k},y_{1}^{k})
  -
  \lambda_{k}(x_{1}^{n},y_{1}^{k}) \right)
  f_{\rvax_{k+1}^{n}|\rvax_{1}^{k}}(x_{k+1}^{n}|x_{1}^{k})dx_{k+1}^{n}
  \\&=
  \int
\left(  \ln K_{k+1}(x_{1}^{n},y_{1}^{k})
-
\ln \breve{F}_{k}(x_{1}^{k},y_{1}^{k})
\right)
  f_{\rvax_{k+1}^{n}|\rvax_{1}^{k}}(x_{k+1}^{n}|x_{1}^{k})dx_{k+1}^{n}
  \\
\iff 
\ln \breve{F}_{k}(x_{1}^{k},y_{1}^{k})
&=
  \int 
\ln K_{k+1}(x_{1}^{n},y_{1}^{k})
  f_{\rvax_{k+1}^{n}|\rvax_{1}^{k}}(x_{k+1}^{n}|x_{1}^{k})dx_{k+1}^{n}
\end{align}
This yields that the recursion~\eqref{subeq_recursion_F} 
takes the form 
\begin{align}
  \breve{F}_{n}(x_{1}^{n},y_{1}^{n})
  &=
  f_{\rvay_{1}^{n}}(y_{1}^{n})
  \\
  \breve{F}_{k}(x_{1}^{k},y_{1}^{k})
  &
  =
  \expo{ \int
\ln \left(\sumover{y_{k+1}}  \expo{-s\rho(x_{k+1},y_{k+1})} \breve{F}_{k+1}(x_{1}^{k+1},y_{1}^{k+1})  \right)
  f_{\rvax_{k+1}^{n}|\rvax_{1}^{k}}(x_{k+1}^{n}|x_{1}^{k})dx_{k+1}^{n}}
  \label{eq:breve_F_k_semifinal}
\end{align}
If $\rvax_{1}^{n}$ is $\kappa$-th order Markovian, then 
$
f_{\rvax_{k}^{n}|\rvax_{1}^{k-1}}(x_{k}^{n}|x_{1}^{k-1})
=
f_{\rvax_{k}^{n}|\rvax_{k-\kappa}^{k-1}}(x_{k}^{n}|x_{k-\kappa}^{k-1})
$,
for all $k=1,\ldots,n$, in which case~\eqref{eq:breve_F_k_semifinal} becomes~\eqref{subeq:recursion_breve_F}. 
Substituting the latter into~\eqref{eq:Fk_breve_Fk} and then in~\eqref{eq:simple_op_F} yields~\eqref{eq:p_y_given_x_and_y_final_op_F}. 

Finally, from~\eqref{eq:p_y_given_x_and_y_final_op_F}, it follows that in a realization of the causal IRDF it must hold that
\begin{align}
  \rvay_{k}
  \longleftrightarrow
  \rvay_{1}^{k-1},\rvax_{k-\kappa+1}^{k}
  \longleftrightarrow
  \rvax_{1}^{k-\kappa}
\end{align}
and that 
\begin{align}
  \rvay_{k}
  \longleftrightarrow
  \rvay_{1}^{k-1},\rvax_{k}
  \longleftrightarrow
  \rvax_{1}^{k-1}
\end{align}  
  does not hold, except for $k=n$.
  This completes the proof.
  \findemo


\end{document}